\documentclass[11 pt,a4paper]{article}
\usepackage[utf8]{inputenc}
\usepackage{amsmath}
\usepackage{amsfonts}
\usepackage{amssymb}
\usepackage{pgfplots}
\pgfplotsset{compat=1.14}
\usepackage{tikz}
\usepgfplotslibrary{fillbetween}
\usetikzlibrary{patterns}
\usepackage{indentfirst}
\usepackage{amsmath}
\usepackage{graphicx}
\usepackage{apalike}

\author{Dr. Jacques Balayla MD, MPH, FRCSC\footnote{To whom correspondence should be addressed: Dr. Jacques Balayla MD, MPH, FRCSC. Quilligan Scholar. e-mail: jacques.balayla@mail.mcgill.ca. Department of Obstetrics and Gynaecology. McGill University, Montreal, Quebec, Canada}}

\title{{{\Large {\textbf{F}ourier \textbf{E}valuation of \textbf{T}racings and \textbf{A}cidosis in \textbf{L}abor: }}
{\Large{The FETAL Technique}}}}
\date{}

\begin{document}
\maketitle  

\begin{abstract}	
Adequate fetal and neonatal development depend upon the presence of a normal acid-base environment during pregnancy and the smooth transition from intra-uterine to extra-uterine life. Current methods to assess fetal pH and acid-base status are invasive and carry significant maternal and fetal risks. Given these limitations, obstetrical care providers developed the electronic fetal monitoring (EFM) system, a non-invasive tool, which evaluates beat-to-beat fetal heart rate (FHR) patterns in order to predict fetal oxygenation status in real-time. Every year, about 85 percent of the approximately 4 million live births in the United States are evaluated using EFM. Unfortunately, though there is ample physiological evidence that FHR patterns are inextricably linked to fetal acid-base status, the use of EFM has not been shown to reliably predict neonatal pH, nor has it reduced the incidence of adverse perinatal outcomes, including long-term neurological morbidity and cerebral palsy (CP). The poor specificity associated with the current interpretation of the EFM therefore leads to a paradox we have henceforth defined as the “Obstetrical Paradox”. In this study, we develop and seek to determine whether a novel, non-invasive method known as the FETAL technique (Fourier Evaluation of Tracings and Acidosis in Labor), which applies the Fourier Transform to EFM tracings and determines the spectral frequency distributions of the FHR, improves the assessment of the fetal pH in real time. We hypothesize that the improvement in the sensitivity and specificity of the EFM with the use of the FETAL technique will lead to a significant reduction in the rate of neonatal hypoxic injury and in the rate of caesarean and assisted vaginal deliveries for suspected fetal distress. The implications of a successful application of the FETAL technique would have paradigm-shifting consequences in the provision of modern obstetrical care.
\end{abstract} 

\newpage

\section{Overview of Fetal Circulatory Physiology}

The fetal-maternal circulation interface is proximate at the placenta, where gas/nutrient exchange between maternal and fetal circulation occurs. Oxygen and nutrients diffuse across the placental membrane from maternal arterial blood and is transported to the fetus via a single large umbilical vein. Following tissue extraction of oxygen and nutrients, fetal blood returns to the placenta via two umbilical arteries. This now deoxygenated blood contains the waste products of fetal metabolism, including carbon dioxide (pCO2), for elimination from maternal circulation via the lungs and kidneys. As such, venous cord blood reflects the combined effect of maternal acid-base status and placental function, whilst arterial cord blood reflects fetal/neonatal acid-base status. The clinical value of cord blood gas analysis in the immediate post-partum period lies in its ability to provide objective evidence of asphyxia at the moment of birth, which is the main determinant of short-term neonatal adaptation to extra-uterine life and long-term neurological status \cite{steer1989interrelationships}. In this regard, cord blood has been shown to be more reliable than routine clinical assessment at birth using the APGAR scoring system \cite{steer1989interrelationships}.

\section{Fetal Acid-Base Status and Metabolism}

The pH, base excess and pCO2 of arterial blood flowing through the umbilical cord provides valuable objective evidence of the metabolic condition of neonates at the moment of birth, including asphyxia. Asphyxia refers to reduced tissue oxygen (hypoxia) of sufficient severity and duration to cause metabolic acidosis. Metabolic acidosis develops when tissue cells are severely depleted of oxygen and aerobic metabolism of glucose is compromised \cite{sykes1982apgar}. Cells must therefore depend on less effective anaerobic pathways that result in reduced ATP (energy) production and accumulation of metabolic acids (principally lactic acid). When normal buffering mechanisms are overwhelmed by this acid influx, the fetal blood pH falls below normal limits. Cord-blood metabolic acidosis – which is characterized by reduced blood pH and decreased base excess (i.e. increased base deficit) – thus implies that sometime during labor, oxygenation of fetal tissues was severely compromised \cite{sykes1982apgar}. 

\section{Cardiotocography - Fetal Heart Rate (FHR)}

Normal human labor is characterized by regular uterine contractions and repeated episodes of transient interruption of fetal oxygenation \cite{parer2006fetal}. Most fetuses tolerate this process well, but some do not. The fetal heart rate (FHR) pattern helps to distinguish the former from the latter as it is an indirect marker of fetal cardiac and central nervous system responses to changes in blood pressure, blood gases, and acid-base status \cite{parer2006fetal}. Because of high interobserver and intraobserver variability in the interpretation of fetal heart rate (FHR) tracings, the American College of Obstetricians and Gynecologists (ACOG), the Society for Maternal-Fetal Medicine (SMFM), and the United States National Institute of Child Health and Human Development (NICHD) convened a workshop to standardize definitions and interpretation of electronic fetal monitoring (EFM) \cite{macones20082008}. The rationale for intrapartum FHR monitoring is that identification of FHR changes potentially associated with inadequate fetal oxygenation may enable timely intervention to reduce the likelihood of hypoxic injury or death \cite{ray2017intrapartum}. Although virtually all obstetric societies advise monitoring the FHR during labor, the benefit of this intervention has not been clearly demonstrated and this position is largely based upon expert opinion and medicolegal precedent \cite{alfirevic2017continuous}.

\section{Fetal Heart Rate (FHR) Intepretation}

Several tracing characteristics are used to interpret the FHR:
\begin{center}
\begin{table}[h]
\center
\begin{tabular}{lccl}
\hline
 & \textbf{Table 1. NICHD - Variables inherent to FHR} \\ \hline
 & Baseline rate  
 \\& Variability 
 \\& Accelerations
 \\& Decelerations
 \\
 \hline
\end{tabular}
\end{table}
\end{center}

As mentioned previously, in 2008, the NICHD set forth recommendations for defining FHR characteristics to improve predictive value of EFM and facilitate evidence-based clinical management of fetal compromise \cite{macones20082008}. The FHR definitions are intended for evaluation of
intrapartum patterns but may be used antepartum. A 3-tier FHR interpretation system was developed, defining a category I FHR tracing as normal, category III as abnormal, and category II as atypical or indeterminate. Category II or III tracings may prompt intervention by the provider, be it intra-uterine resuscitation and/or prompt delivery.

\begin{center}
\begin{table}[h]
\centering
\begin{tabular}{lccl}
\hline
 & \textbf{Table 2. NICHD - FHR Classification} \\ \hline
 & Category I  - Normal
 \\& Category II - Atypical
 \\& Category III - Abnormal
 \\
 \hline
\end{tabular}
\end{table}
\end{center}

\section{Fetal Distress}

Clinically, it is generally preferable to describe specific signs of suspected fetal compromise in lieu of simply providing an all-encompassing diagnosis of fetal distress \cite{parer1990fetal}. These include: 
\begin{itemize}
  \item[$\cdot$] Decreased fetal movements felt by the mother
  \item[$\cdot$] Meconium in the amniotic fluid 
  \item [$\cdot$]Non-reassuring patterns seen on cardiotocography such as:
\begin{itemize}
  \item[$\cdot$] Fetal tachycardia and bradycardia
  \item[$\cdot$] Decreased variability in the fetal heart rate
  \item[$\cdot$] Late decelerations
\end{itemize}
\end{itemize}

Likewise, biochemical signs, assessed by fetal scalp blood sampling can be indicative of fetal distress. In particular, a low pH and elevated blood lactacte levels indicates fetal metabolic acidosis \cite{sykes1982apgar}. Some of these signs are more reliable predictors of fetal compromise than others. For example, cardiotocography can give high false positive rates, even when interpreted by highly experienced medical personnel \cite{alfirevic2017continuous}. Metabolic acidosis is a more reliable predictor, but is not always available.

\section{Problems with the FHR interpretation}
Over the last 30 years, little progress has been made in the screening and diagnosis of fetal acidosis. Though some headway was made with the widespread introduction of electronic fetal monitoring (EFM) in labour, the presence of a marked pattern of late decelerations, a hallmark traditionally associated with poor neonatal outcomes, has been shown to predict neonatal acidosis in less than 50 percent of cases \cite{balayla2019use}. Initially, EFM was introduced into clinical practice without appropriate studies on its validity (relationship of FHR patterns to fetal outcome), reliability (intra- and inter-observer variability), and causal relationship to outcome (ability of intervention to avoid metabolic acidemia). Because of the demonstrable high inter-observer and intra-observer variability in the interpretation of fetal heart rate (FHR) tracings \cite{balayla2019use}, and despite the NICHD management guidelines \cite{macones20082008}, EFM has not substantially changed the incidence of neonatal hypoxia, hypoxic ischemic encephalopathy, neonatal academia, cerebral palsy, or neurodevelopmental impairment. Whereas some evidence suggests that intrapartum fetal monitoring is associated with a reduction in intrapartum death and neonatal seizures, a reduction in long-term neurologic disability – which is primarily related to oxygen status at birth – has not been demonstrated \cite{alfirevic2017continuous}. Though EFM has moderate sensitivity, it has a low specificity for detecting fetal hypoxia/asphyxia. As a consequence, the high false-positive rate of EFM for predicting adverse neonatal outcomes has increased the rates caesarean delivery to an average of 33 percent in North America since the 1970s, with elevated rates of associated maternal and neonatal morbidity \cite{alfirevic2017continuous}. 
\\
\

Though the development of a systematic and objective method of analysis and diagnosis in EFM was theorized to overcome the high inter-observer variability in its interpretation, several large randomized controlled trials (RCTs), including the international INFANT trial, have not demonstrated improvements in neonatal outcomes when computerized and AI systems were used for the quantitative and qualitative analysis of EFMs \cite{balayla2019use}. As we work toward realizing the full potential benefits of EFM, finding the best assessment strategies to evaluate fetal pH in real-time remains a key goal in Obstetrics. 

\section{The Obstetrical Paradox} 
It has been well established that fetal acid-base status is inextricably linked to fetal heart rate patterns  \cite{reddy2009antepartum}. Since the EFM has not been shown to reliably predict neonatal pH status or reduce adverse outcomes, one fundamental conundrum, hereby termed the ``Obstetrical Paradox", arises: if the fetal heart rhythm is inextricably linked to fetal acid-base status, why are neonatal outcomes not improved when tracing changes are detected and acted upon?  \cite{balayla2019use}. Outside of human error and delays in intervention, a potential solution to the obstetrical paradox may be related to the possibility that the modern interpretation of EFM is incomplete - that other visible tracing characteristics we do not presently account for are actually predictive of fetal pH and lead to better outcomes when acted upon. A second possibility is that perhaps there is merely an association and not a causal pathway between changes in pH and EFM fluctuations. Such explanation would imply that there is no direct FHR-pH relationship. Finally, it may be the case that the key to predicting fetal pH lies embedded within EFM tracings but beyond their traditional clinical interpretation as set by the NICHD \cite{macones20082008}. By applying the Fourier Transformation and determining the frequency distribution of EFM tracings, the FETAL technique seeks to delve into the EFMs’ fundamental build-up components to test the latter hypothesis in a clinical setting. 

\newpage
\textbf{Figure 1. The Obstetrical Paradox and its potential solutions}
\begin{figure}[h!]
\centering
\includegraphics[scale=0.112]{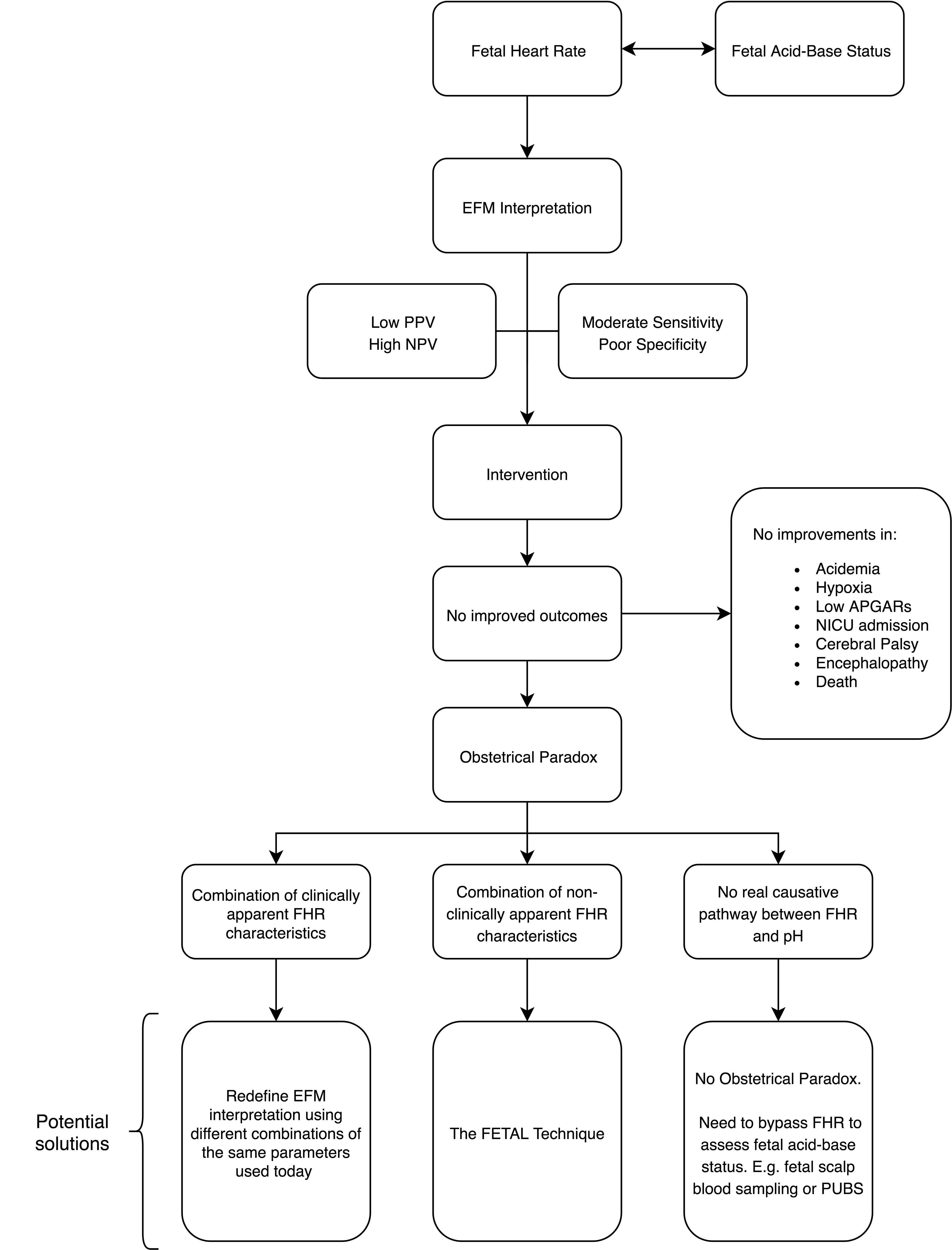}
\end{figure}
\section{The Fourier Transform}
The Fourier transform (FT) is a mathematical tool which decomposes either a periodic or non-periodic continuous function of time, also known as a signal, into individual sinusoidal waves of specific amplitude, frequency, and phase shift \cite{stein2011fourier}. The inverse Fourier transform mathematically synthesizes the original time-dependent function from these sinusoidal waves and their frequency domain representation.
\\
\

As such, the Fourier transform can be used to yield the frequency spectrum of the original signal. Each sinusoidal component is itself a time-dependent, complex-valued function of frequency, whose magnitude, or modulus, represents the amount of that frequency present in the original function. Likewise, its argument is the phase offset of the basic sinusoid for that frequency. In practical terms, this means that the Fourier transformation of a signal yields a list of individual sinusoidal functions with specific frequencies and magnitudes, which when added, reveal the original function anew.
\\
\

The Fourier transform is not limited to functions of time, but the domain of the original function is commonly referred to as the \textit{time domain}.  Concretely, this means that any linear time-invariant system, such as the fetal heart rate signal, can be expressed relatively simply as the sum of frequencies of specific sine and cosine waves. When both the function and its Fourier transform are replaced with discretized counterparts, it is called the discrete Fourier transform (DFT). The DFT has become a mainstay of numerical computing in part because of a very fast algorithm for computing it, called the Fast Fourier Transform (FFT), which was known to Gauss (1805) and was brought to light in its current form by Cooley and Tukey. 
\\
\

There are several common conventions for defining the Fourier transform of an integrable function $f:\mathbb {R} \to \mathbb {C}$. For clarity, we will define the transform equation as $X(\omega)$ since we'll be addressing frequency in Hertz (e.g. if time is measured in seconds, then the frequency is in Hertz):

\begin{large}
\begin{equation}
{{X}}(\omega) = {\int _{ -\infty }^{ \infty }x(t)e^{-2\pi j\omega t}dt}
\end{equation}
\end{large}

When the independent variable t represents time, the transform variable $\omega$ represents frequency. Under suitable conditions, x(t)  is determined by X$(\omega)$ via the inverse transform:

\begin{large}
\begin{equation}
{x(t) = {\int _{ -\infty }^{ \infty }{{X}}(\omega)}e^{2\pi j\omega t}d\omega}
\end{equation}
\end{large}

To render the above equation more manageable and to help integrate it, we consider the Taylor series expansion:

\begin{large}
\begin{equation}
f(x) = \sum_{n=0}^{\infty} \frac{f^{(n)}(x_0)}{n!}(x-x_0)^n
\end{equation}
\end{large}

Since $e^x$ is its own derivative, the Taylor series expansion for $f(x) = e^x$ is rather simple:

\begin{large}
\begin{equation}
e^x = \sum_{n=0}^{\infty} \frac{x^n}{n!} = 1 + x + \frac{x^2}{2!} + \frac{x^3}{3!}...
\end{equation}
\end{large}

When incorporating the complex plane into the definition, we obtain:

\begin{large}
\begin{equation}
e^{i\theta} = \sum_{n=0}^{\infty} \frac{(i\theta)^n}{n!} = 1 + i\theta - \frac{\theta^2}{2!} - \frac{i\theta^3}{3!}...
\end{equation}
\end{large}

so that for the set of all positive integers $p$:

\begin{large}
\begin{equation}
cos(\theta)= \sum_{n=2p}^{\infty} \frac{(-1)^\frac{n}{2}}{n!}\theta^n
\end{equation}
\end{large}

and 

\begin{large}
\begin{equation}
sin(\theta)=\sum_{n=2p-1}^{\infty} \frac{(-1)^\frac{n-1}{2}}{n!}\theta^n
\end{equation}
\end{large}

As such, all even order terms are real and all odd order terms are imaginary, thereby yielding Euler's formula:

\begin{large}
\begin{equation}
e^{i\theta} = cos(\theta) + i sin(\theta)
\end{equation}
\end{large}

Re-writing sines and cosines as complex exponentials makes it necessary for the Fourier coefficients to be complex valued. The usual interpretation of this complex number is that it gives both the amplitude (or size) of the wave present in the function and the phase (or the initial angle) of the wave. These complex exponentials sometimes contain \textit{negative} frequencies. If $\theta$ is measured in seconds, then the waves $e^{2i\pi\theta}$ and $e^{-2i\pi\theta}$ both complete one cycle per second, but they represent different frequencies in the transform. Given a trajectory, the Fourier transform yields individal waves, each of which has a strength, a delay and a speed \cite{stein2011fourier}. 
\\
\

The trajectory is processed through a set of filters, whereby each filter yields a cycle and the remainder of the trajectory filters are independent. Otherwise stated, each one catches a different part of the trajectory until there are enough filters to catch all of the trajectory, ie, the last filter leaves no trajectory remainder. The result cycles can be combined linearly, giving the same results no matter the mixing order. In order to accomplish the latter, two algorithms exist:
the Discrete Fourier Transform (DFT), which requires $O(n^2)$ operations (for n samples), and the Fast Fourier Transform (FFT) which requires $O(n.log(n))$ operations \cite{stein2011fourier}.

\subsection{Frequency Domain}

The frequency domain refers to the analysis of mathematical functions or signals with respect to frequency, rather than time. Put simply, a time-domain graph shows how a signal changes over time, whereas a frequency-domain graph shows how much of the signal lies within each given frequency band over a range of frequencies. A frequency-domain representation can also include information on the phase shift that must be applied to each sinusoid in order to be able to recombine the frequency components to recover the original time signal. The `spectrum' of frequency components is the frequency-domain representation of the signal. The inverse Fourier transform converts the frequency-domain function back to the time function. A spectrum analyzer is a tool commonly used to visualize electronic signals in the frequency domain. Some specialized signal processing techniques use transforms that result in a joint time–frequency domain, with the instantaneous frequency being a key link between the time domain and the frequency domain.
Homogeneity means that a change in
amplitude in one domain produces an identical change in amplitude in the other
domain. This should make intuitive sense: when the amplitude of a time
domain waveform is changed, the amplitude of the sine and cosine waves
making up that waveform must also change by an equal amount.

\subsection{Properties of the FT}
The Fourier Transform is linear, that is, it possesses the properties of homogeneity and additivity. Homogeneity means that a change in amplitude in one domain produces an identical change in amplitude in the other domain. Additivity of the Fourier transform means that addition in one domain corresponds to addition in the other domain.

\begin{large}
\begin{equation}
{\hat {f}}(\xi) = {\int _{ -\infty }^{ \infty }f(x)e^{-2\pi i\xi x}dx}\leftrightarrow{k\hat {f}}(\xi)= k{\int _{ -\infty }^{ \infty }f(x)e^{-2\pi i\xi x}dx}
\end{equation}
\end{large}

\subsection{The Discrete Fourier Transformation}
The integral range in equation (1) goes from $-\infty$ to $\infty$. Since the fetal heart rate signal is not periodic nor perpetual, this infinite integral will not be a useful tool for this purpose. For the purposes of our analysis, we must employ the $Discrete$ Fourier Transform (DFT). The discrete Fourier transform (DFT) converts a finite sequence of equally-spaced samples of a time function into a same-length, equally-spaced complex-valued function of frequency via an algorithm known as the Fast Fourier Transformation (FFT) \cite{walker2017fast}. The interval at which the DFT is sampled is the reciprocal of the duration of the input sequence.
\\
\

The DFT formula is defined as:
\begin{large}
\begin{equation}
{X_k = \sum_{n=0}^{N-1} x_n\cdot e^{-\frac {i 2\pi}{N}kn}}
\end{equation}
\end{large}
\
where $N$ refers to the total number of samples, $n$ refers to the `nth' sample, $x_n$ refers to the function value at the $nth$ sample, and $k$ is the frequency attributed to the ensuing sine wave at the `nth' sample.
\\
\

The analogue to the continuous form in the case of the discrete form is as follows:
\\
\
\begin{large}
\begin{equation}
\omega={\frac {k}{N}}
\end{equation}
\end{large}
and
\begin{large}
\begin{equation}
n = t
\end{equation}
\end{large}
This relationship can be re-written as:
\begin{large}
\begin{equation}
{X_k = \sum_{n=0}^{N-1} x_n\cdot e^{-\frac {i 2\pi}{N}kn}}=x_0e^{-b_0i}+x_1e^{-b_1i}+...+x_ne^{-b_{N-1}i}
\end{equation}
\end{large}
\\
\
where $b_N$ = $2\pi {\frac {kn}{N}}$, $X_k$ is the `kth' frequency bin and $x_n$ is the `nth' sample.
\\
\

By considering Euler's Formula in (8) and replacing $\theta$ with ${b_{0}}$, we can expand equation (13) as follows:
\\
\
\begin{small}
\begin{equation}
{X_k = x_0[cos(-b_0) + isin (-b_0)] + ... + x_n[cos(-b_{N-1}) + isin (-b_{N-1})]}
\end{equation}
\end{small}
We can further express the $X_k$ relationship as:
\begin{large}
\begin{equation}
X_k = A_k + B_ki
\end{equation}
\end{large}
\begin{center} 
where $A_k$ = $x_0 cos(-b_0)$ and $B_k$ = $x_0 sin(-b_0)$
\end{center} 
With this notation, we can represent $X_k$ in the complex plane in vector form:
\\
\
\begin{center}

\begin{tikzpicture}
 
	\begin{axis}[
    xlabel = $Real$,
	ylabel = {$Imaginary(i)$},    
     ymin=0, ymax=5,
     xmin=0, xmax=5,
    legend pos = south east,
     ymajorgrids=false,
    grid style=dashed,
    width=7cm,
    height=7cm,
     ]
	\end{axis}
  \draw[->, ultra thick, blue,  arrows={-latex}]  (0,0) -- (4,3) node[sloped,midway,above=-0.1cm] {$\mathsf{X_k}$};
\draw[->, ultra thick, red, dashed, arrows={-latex}]  (0,0) -- (4,0) node[sloped,midway,above=-0.1cm] {$\mathsf{A_k}$};
\draw[->, ultra thick, red, dashed, arrows={-latex}]  (4,0) -- (4,3) node[sloped,midway,above=-0.1cm] {$\mathsf{B_ki}$};
\end{tikzpicture}
\end{center}

\begin{center} 
Note each component of $X_k$ in red, in each axis.
\end{center}

To calculate the magnitude of the vector, which will give us the magnitude of the frequency component, we simply apply Pythagoras' theorem, as follows:
\begin{equation}
{\vert{X_k}\vert = \sqrt{A_k^2 + B_k^2}}
\end{equation}
\

Likewise, to calculate phase shift $\theta$ for the sinusoidal wave with that particular frequency we need to use the $arctan$ function:
\begin{equation}
\theta=\arctan{(\frac{B_k}{A_k})}
\end{equation}
\

An N-point DFT is expressed as the multiplication   X=Wx, where x is the original input signal, W is the N-by-N square DFT matrix, and X is the DFT of the signal. The transformation matrix W can be defined as W = $\left(\frac{\omega^{ik}}{{\sqrt{N}}}\right)_{i,k=0,\ldots,N-1}$, or otherwise stated:
\[
W = \frac{1}{\sqrt{N}}
  \begin{bmatrix}
    1&1&1&1&\cdots &1 \\
1&\omega&\omega^2&\omega^3&\cdots&\omega^{N-1} \\
1&\omega^2&\omega^4&\omega^6&\cdots&\omega^{2(N-1)}\\ 1&\omega^3&\omega^6&\omega^9&\cdots&\omega^{3(N-1)}\\
\vdots&\vdots&\vdots&\vdots&\ddots&\vdots\\
1&\omega^{N-1}&\omega^{2(N-1)}&\omega^{3(N-1)}&\cdots&\omega^{(N-1)(N-1)}
  \end{bmatrix}
\]
\\
\
where $\omega = e^{-2\pi i/N}$ is a primitive Nth root of unity in which $i^{2}=-1$.

\section{Fourier Transform in FHR analysis}
The utility of the Fourier transformation in the analysis of FHR stems from the spectrum analysis of the frequency-domain, which is aided by a fundamental theorem in probabilistic theory, namely, the law of large numbers \cite{hsu1947complete}.
\subsection{The Law of Large Numbers}

Two different versions of the law of large numbers exist; they are called the strong law of large numbers, and the weak law of large numbers \cite{hsu1947complete}. Stated for the case where X1, X2, ... is an infinite sequence of independent and identically distributed Lebesgue integrable random variables with expected value E(X1) = E(X2) = ...= $\mu$, both versions of the law state that – with virtual certainty – the sample average:
\\
\
\begin{equation}
\overline{X}_n=\frac1n(X_1+\cdots+X_n)
\end{equation}
\\
\
converges to the expected value
\\
\
\begin{equation}
\overline{X}_n \, \to \, \mu \ for \ n \to \infty
\end{equation}

Otherwise stated, as $\ n \to \infty$
\begin{equation}
P(\lim\overline{X}_n = \mu) =1
\end{equation}

Though a frequency-domain spectrum is not an individual variable, the law of large numbers still applies regarding its range. If we assume that a normal fetal physiologic state exists from fetus to fetus, their individual FHR frequency spectrum range, on average, will be composed of similar frequencies. By sampling an ever-increasing number of FHRs with normal pH at birth, we can determine the expected frequency spectrum range in a normal fetus and its 95 percent confidence interval. While not a theoretical pre-test calculation, this established range can be used as the expected theoretical value against which future assessments can be compared. This allows us to contrast between a fetus with normal frequency parameters and one without. Whether an abnormal frequency spectrum is associated with an abnormal pH is what this technology seeks to determine.

\subsection{The Kolmogorov–Smirnov test (K–S test)}
As mentioned previously, the law of large numbers will allow us to establish a reference range for the normal frequency domain of the FHR of a term fetus. Once an expected, reference curve is created, we can use it to compare any individual fourier-transformed FHR strip against it to determine whether its shape is significantly different with the use of the Kolmogorov–Smirnov test. The Kolmogorov–Smirnov test, or K-S test for short, is a nonparametric test of the equality of continuous, one-dimensional probability distributions that can be used to compare a sample with a reference probability distribution (one-sample K–S test), or to compare two individual samples (two-sample K–S test). With a one-sample K-S test, a distribution is tested for normality. That is, it is tested against a normal Gaussian distribution of mean $\mu$ and variance $\sigma^2$ \cite{lopes2007two}. In the case of the FETAL technique, we will use the two-sample K-S test to compare a given Fourier-transformed FHR strip to the reference curve. Simply stated, the Kolmogorov–Smirnov statistic quantifies a distance between the empirical distribution function of the sample and the cumulative distribution function of the reference distribution.
\\
\

The empirical distribution function $F_n$ for $n$ independent and identically distributed ordered observations $X_i$ is defined as:

\begin{equation}
F_n(x)={1 \over n}\sum_{i=1}^n I_{[-\infty,x]}(X_i)
\end{equation}
\

where $I_{[-\infty,x]}(X_i)$ is the indicator function, equal to 1 if $X_i$ $\leq$ $x$ and equal to 0 otherwise.
\\
\

The Kolmogorov–Smirnov statistic $D_n$ for a given cumulative distribution function $F(x)$ is given by the equation:
\begin{equation}
D_n= \sup_x |F_n(x)-F(x)|
\end{equation}
\
where $\sup_x$ is the supremum of the set of distances between both distributions. When using a two-sample K-S test, the Kolmogorov–Smirnov statistic can be defined as:
\begin{equation}
D_{n,m}=\sup_x |F_{1,n}(x)-F_{2,m}(x)|
\end{equation}

For large samples, the null hypothesis is rejected at level $\alpha$, the probability of rejecting the null hypothesis when the null hypothesis is true, if:

\begin{equation}
D_{n,m}>c(\alpha)\sqrt{\frac{n + m}{n m}}
\end{equation}

Where  $n$ and $m$ m are the sizes of first and second sample respectively. The value of $c({\alpha})$ is given by the equation below:

\begin{equation}
c(\alpha)=\sqrt{-\frac{1}{2}\ln\alpha}
\end{equation}
\subsection{Dimensionality of the Fourier Transform}
When the time-domain signal is transformed into a frequency-domain via the fourier transform, the axes change. While in the time-domain the independent axis is $time$, the independent axis in the frequency domain is that of $frequency$ in Hertz (Hz), and thus, it is time-independent. In both cases the dependent, $y$ axis is the amplitude or intensity of the signal. A theoretical example of a FHR strip is presented here:
\begin{center}

\begin{tikzpicture}
	\begin{axis}[
    axis lines = center,
    xlabel = {$time (sec)$},
	ylabel = {$bpm$},    
     ymin=90, ymax=180,
    legend pos = north east,
     ymajorgrids=true,
     xmajorgrids=true,
     xminorgrids=true,
    grid style=dashed,
    width=10cm,
    height=5cm,
     ]
	\addplot [
	domain= 0:500,
	color= blue,
	]
	coordinates {
    (0,135)(1,137)(2,133)(4,134)(7,142)(10,149)(20,138)(30,143)(40,144)(60,146)(80,133)(100,135)(102,133)(104,151)(110,150)(120,155)(130,153)(140,130)(150,139)
    };
    \addplot+ [
dashed,
domain= 0:150,	
color = red,
mark size = 0pt
 ]
	{110};
	    \addplot+ [
dashed,
domain= 0:150,	
color = red,
mark size = 0pt
 ]
	{160};
	\end{axis}
\end{tikzpicture}
\end{center}
Likewise, a computerized power spectrum analysis of the frequency domain by Romano et al.  \cite{romano2016frequency} is seen below. Note its bimodal/bi-band distribution in the low and high-frequency range:
\begin{figure}[h!]
\centering
\includegraphics[scale=0.26]{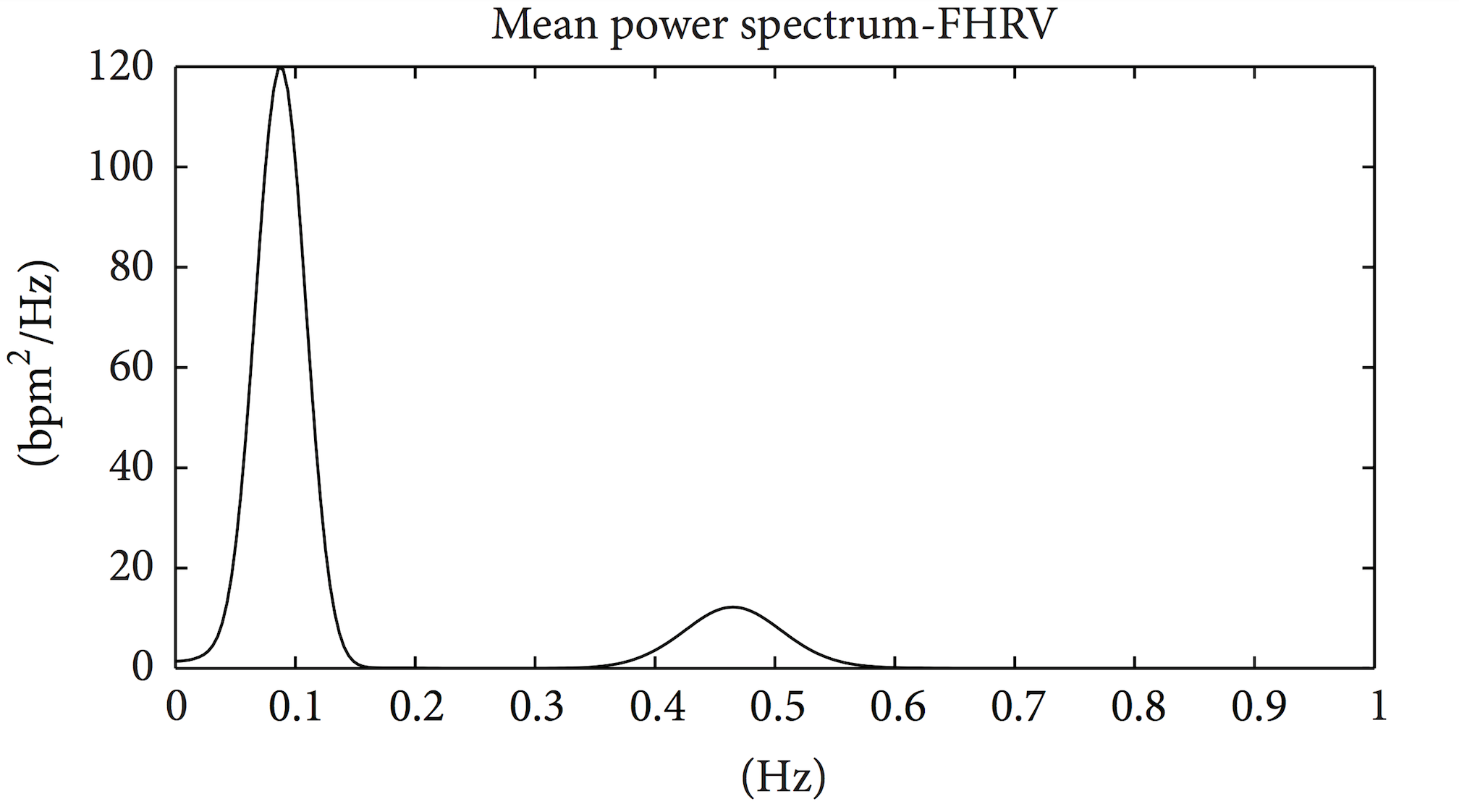}
\end{figure}
\subsection{Physiological basis for the FETAL technique}
Heart rate variability investigations started in Obstetrics, with the observation that changes in FHR variability precede changes in actual heart rate in cases of intrauterine asphyxia. Indeed, the heart rate depends on the sinus node’s intrinsic rate and sympathetic-parasympathetic nervous tone balance, which themselves are directly dependent on present oxygenation status \cite{doret2015fractal}. Power spectral analysis and Fourier transformation of FHR variability has shown that sympathetic and parasympathetic nervous activities make frequency-specific contributions to the heart rate power spectrum, and that renin-angiotensin system activity strongly modulates the amplitude of the spectral peak located at 0.04 Hz. Specifically in the non-anomalous term fetus, heart rate variability estimated by the high-frequency (HF) bands between 0.35–0.45 Hz reflects FHR control by the parasympathetic tone and the low-frequency (LF) bands between 0.03–0.15 Hz reflects the sympathetic tone. Studies of heart rate fluctuation based on frequency analysis have been carried out in animal models. However, their clinical application in humans has not been studied as of yet. Previous studies have found that the shift of autonomic balance is related to the redistribution of the power between the LF and HF bands, and that normalized power units or LF/HF are effective methods of determining the shift of autonomic balance – the mechanism presumed to be at the helm of fetal acid-base status. A higher proportion of LF bands evokes a predominantly sympathetic environment, which is indicative of a stress response. What threshold of that stress response is indicative of fetal hypoxemia is unknown. By studying specific low- and high-frequency patterns of Fourier-transformed domains, the FETAL Study seeks to determine whether fetal acid-base status evaluations can be made non-invasively and in real-time. Examples by Doret et al. of Fast Fourier Transformations, power spectrum densities (PSD) and log-converted frequency wavelet distributions superimposed on the PSD of the EFM are illustrated here \cite{doret2015fractal}:
\\
\
\begin{figure}[h!] 
\centering
\includegraphics[scale=0.183]{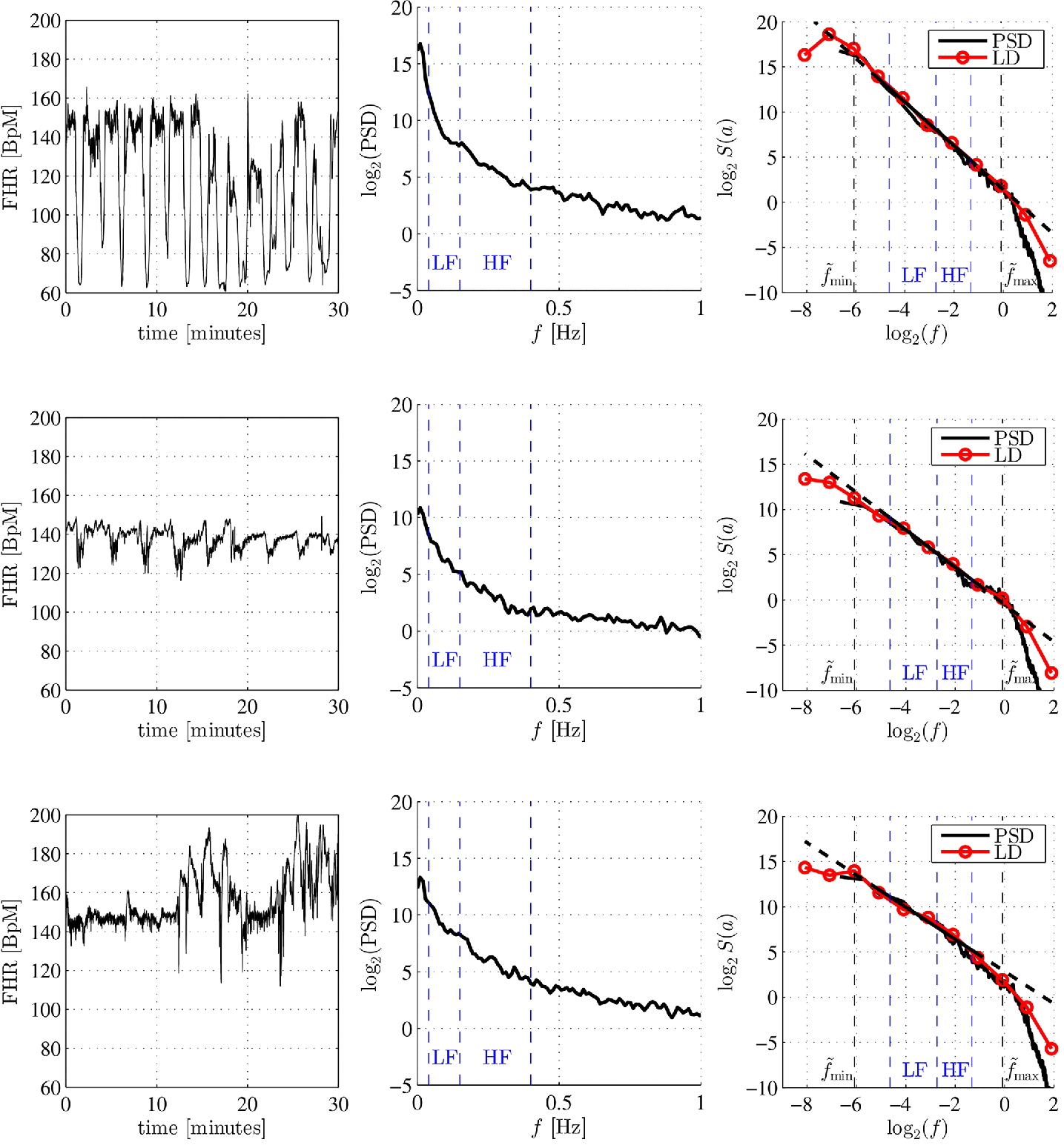}
\end{figure}

\subsection{Procedural Steps for the FETAL Technique}
In order to develop this technology for clinical use, we first need to obtain a reference range of Fourier-transformed frequencies in a normal newborn population. Based on statistical inference, we would need a minimum of 500 tracings of newborns with normal umbilical artery pH levels between 7.15 - 7.35, in the context of normal base excess and lactate levels. This would allow us to determine a normal reference range in the frequency domain against which we can compare future frequency distributions. Using the discrete Fourier transformation and the Kolmogorov–Smirnov test in real-time at the bedside, we can assess whether a frequency distribution significantly deviates from the reference range developed. Such deviation would be a warning sign prompting an intervention from the provider to reduce rates of abnormal pH. Likewise, distributions consistent with the reference range are re-assuring and need not prompt immediate action. Rates of Caesarean and assisted-vaginal delivery are expected to decrease as a consequence.
\section{Conclusion}
Adequate fetal and neonatal development depend upon the presence of a normal acid-base environment during pregnancy and the smooth transition from intra-uterine to extra-uterine life. Current screening methods using the EFM are inadequate to predict and prevent injury, likely as a consequence of the poor inter-rater reliability observed amongst healthcare professionals.  While we do not expect the FETAL technique, should it be successful, to completely overcome the issues pertaining to inter-rater reliability as well as the intricacies of clinical decision making, it is conceivable that outcomes would be improved relative to their state today. Indeed, we hypothesize that the improvement in the sensitivity and specificity of the EFM with the use of the FETAL technique will lead to a significant reduction in the rate of neonatal hypoxic injury and in the rate of caesarean and assisted vaginal deliveries for suspected fetal distress. The implications of a successful application of the FETAL technique would have paradigm-shifting consequences in the provision of modern obstetrical care.
\newpage

\bibliographystyle{apalike}
\bibliography{references}

\end{document}